\documentclass[reprint,amsmath,amssymb,aps,pre]{revtex4-2}

\usepackage{amsmath}
\usepackage{amssymb}
\usepackage{amsfonts}
\usepackage{appendix}
\usepackage{graphicx}
\usepackage{comment}
\usepackage{hyperref}
\usepackage{xcolor}
\usepackage{lineno}

\begin{document}

\title{Urban scaling and the dynamics of individual cities}

\author{Ulysse Marquis$^{1,2}$}
\email{ulyssepierre.marquis@unitn.it}
\author{Marc Barthelemy$^{3,4,5}$}
\email{marc.barthelemy@ipht.fr}
\affiliation{$^1$ Fondazione Bruno Kessler, Via Sommarive 18, 38123 Povo (TN), Italy}
\affiliation{$^2$ Department of Mathematics, University of Trento, Via Sommarive 14, 38123 Povo (TN), Italy}
\affiliation{$^3$ Universit\'e Paris-Saclay, CNRS, CEA, Institut de Physique Th\'eorique, 91191, Gif-sur-Yvette, France}
\affiliation{$^4$ Centre d’Analyse et de Math\'ematique Sociales (CNRS/EHESS) Paris, France}
\affiliation{$^5$ Complexity Science Hub, Vienna, Austria}

\date{\today}

\begin{abstract}

Cities are often compared through scaling laws that relate population size to aggregate urban quantities such as infrastructure, socioeconomic activity, or environmental impacts. These laws are influential because the scaling exponent describing how an urban quantity varies with population is often interpreted as indicating what happens when a city grows, with implications for urban theory, planning, and policy. Here, we show that this interpretation is generally misleading. A scaling relationship obtained by comparing many cities at one point in time does not, in general, describe the trajectory of any individual city. Instead, it reflects a statistical pattern across cities with different histories, constraints, institutions, and growth paths. Relationships across cities can therefore suggest that an urban quantity increases either more slowly or more rapidly than population, even when individual cities follow much simpler dynamics, as we show for the relation between urban area and population. Cross-sectional scaling laws should therefore not be used to infer growth mechanisms or guide policy for individual cities: they describe statistical relationships across cities, rather than the temporal dynamics of individual cities.


\end{abstract}

\maketitle


\section{Introduction}

Most cities are growing, and this growth affects many urban quantities at once:
built area, housing, roads, transport systems, services, emissions, wages,
innovation, crime, health, and many others. A central quantitative question is
therefore how these quantities change as city population increases. To address
this issue, the urban scaling framework was introduced
\cite{Zipf1949LeastEffort,NarollBertalanffy1956Allometry,nordbeck1971allometric,woldenberg1973allometric,longley1991size,pumain2004scaling,Kuhnert2006,bettencourt2007growth,chen2010characterizing}.
In its simplest form, it relates an aggregate urban quantity \(Y\) to population
\(P\), used as a proxy for city size \cite{batty2008}, through
\begin{align}
\label{eq:scaling}
Y \sim P^\beta \,.
\end{align}
In practice, the exponent \(\beta\) is usually estimated from datasets
containing many cities observed at a fixed time. The resulting relation is
therefore cross-sectional: it describes an ensemble of cities at one moment, not
the temporal evolution of a single city.

This framework has generated a large literature reporting apparent scaling laws
for urban heat islands~\cite{oke1973}, economic
productivity~\cite{Sveikauskas1975,pumain2006evolutionary},
crime~\cite{glaeser1999,puttock2025larger}, infrastructure
networks~\cite{Kuhnert2006,Samaniego2008},
innovation~\cite{bettencourt2007invention,Arbesman2009}, pollution and
CO\(_2\) emissions~\cite{Louf_2014,glaeser2010greenness,oliveira2014green,rybski2017cities,fragkias2013does},
building heights~\cite{schlapfer2015urbanskylinesbuildingheights}, road
casualties~\cite{cabrera2020}, and depression rates~\cite{stier2021}. Three
regimes are usually distinguished. When \(\beta<1\), the per-capita value of
\(Y\) decreases with city size, which is often interpreted as an economy of
scale. When \(\beta>1\), \(Y\) grows more than proportionally with population,
suggesting increased returns or amplification effects. The case \(\beta=1\)
corresponds to linear scaling.

These interpretations matter because scaling exponents are often treated not
only as descriptive quantities, but also as indicators of mechanisms. For
example, sublinear scaling of CO\(_2\) emissions can be read as evidence that
larger or denser cities are intrinsically more carbon-efficient
\cite{glaeser2010greenness,oliveira2014green,fragkias2013does}, while
superlinear scaling of innovation or wages is often taken as evidence for
agglomeration benefits. More generally, urban scaling exponents are frequently
interpreted as revealing generic mechanisms of urban growth, efficiency, and
increasing returns.

This interpretation is deeply problematic. First, studies such as
Eq.~\eqref{eq:scaling} are shaped by data availability: cross-sectional
snapshots of systems of cities are common, whereas long time series for
individual cities are scarcer. This imbalance has encouraged the view that
scaling exponents capture a fundamental property of urban systems, although
their connection to individual trajectories is far from clear
\cite{xu2019crosssectionalurbanscalingfails,pumain2010ergodicity,ribeiro2020relation,bettencourt2020interpretation}.
Second, the measured exponent depends on how cities are defined and how data are
aggregated \cite{Louf2014,Arcaute2015}. Third, small deviations from unity do not
necessarily imply nonlinear urban dynamics
\cite{Shalizi2011,Leitao2016,barthelemy2019tomography}.

The central question is therefore: \emph{what does the exponent \(\beta\)
measured across cities actually mean for an individual city?} Previous work has
already exposed this difficulty. For congestion-induced delays, aggregating data
across cities and years yields an apparent nonlinear scaling, whereas individual
city trajectories display strong path dependence and no common law
\cite{depersin2018global}. Other studies have shown clear mismatches between
cross-sectional regularities and city-level dynamics
\cite{ribeiro2020relation,xu2019crosssectionalurbanscalingfails}. Ribeiro et
al.~\cite{ribeiro2020relation} found city-specific longitudinal exponents for
Brazilian cities, while Bettencourt et al.~\cite{bettencourt2020interpretation}
emphasized that temporal exponents can become very large when urban quantities
contain strong temporal trends or are better described by forms other than power
laws (see Supplementary section I for additional discussions on \cite{ribeiro2020relation,bettencourt2020interpretation}). Similar conclusions arise from micro-level wage data, where superlinear
cross-sectional scaling coexists with highly heterogeneous city trajectories
\cite{keuschnigg2019scaling}. Historical path dependence and technological
change further complicate the interpretation of relations measured across
systems of cities \cite{connor2024frontier,connor2025big}.

Pumain~\cite{pumain2010ergodicity} argued that giving a dynamical meaning to
cross-sectional scaling implicitly assumes a form of ergodicity in systems of
cities. In this context, ergodicity means identifying the temporal evolution of
one city with the ensemble pattern observed across many cities at one date. This
requires the strong assumption that cities have comparable histories and can be
viewed, at least approximately, as scaled-up versions of one another.

These observations point to a fundamental issue: Eq.~\eqref{eq:scaling} does
not imply that cities evolve along the same trajectories in the \((Y,P)\) plane.
It only constrains their relative positions at a given time. Even if every city
follows a well-defined relation between \(Y\) and \(P\) over time, the exponent
measured across cities need not equal the growth exponent of any individual
city. Rather, as we show below, it is an emergent statistical quantity shaped by
heterogeneous trajectories, historical contingencies, inter-city correlations,
and the distribution of city sizes.

In this paper, we ask what the exponent measured across cities tells us about
urban growth. We show that, in general, it does not describe the trajectory of
any individual city. Instead, it combines how cities change over time, how they
differ from one another, and how city size correlates with city-specific
characteristics. This decomposition identifies the restrictive conditions under
which a scaling exponent can be interpreted as a genuine law of urban growth.
More generally, urban scaling relations should be understood as statistical
patterns emerging from a diverse system of cities, not as rules describing the
growth of individual cities.

\section{Results}

We compare two perspectives on urban change: following the same cities through time, and comparing many cities at one date. These perspectives do not necessarily agree. Here, we start from individual trajectories and interpret the cross-sectional pattern as the outcome of many heterogeneous histories observed in a single snapshot. The main question is therefore: what information does a cross-sectional exponent provide about cities, and their evolution? A more detailed discussion of previous approaches, and of the distinction between temporal and transversal scaling, is provided in section I the Supplementary Material.

\subsection{Cross-sectional versus longitudinal scaling}

We first clarify the distinction between transversal or cross-sectional scaling and longitudinal dynamics. 
Consider \(N\) cities at time \(t\), with populations
\(P_1(t),\ldots,P_N(t)\) and urban quantities
\(Y_1(t),\ldots,Y_N(t)\). Cross-sectional, or transversal, scaling describes how
\(Y\) varies across cities at fixed \(t\):
\begin{equation}
Y_i(t) \approx Y_0(t)\,P_i(t)^{\beta_T(t)} ,\qquad \forall i,\;\mathrm{fixed}\;t
\label{eq:beta0}
\end{equation}
where \(\beta_T(t)\) is the transversal exponent and \(Y_0(t)\) a
time-dependent prefactor. In log--log coordinates, \(\beta_T(t)\) is usually
estimated as the slope of a regression across cities at fixed time.

By contrast, longitudinal dynamics describe the evolution of \(Y_i\) within one
city \(i\) as its population changes. In full generality,
\begin{align}
Y_i(t) = F_i\!\big(P_i(t),t\big) \qquad \forall t,\; \mathrm{fixed}\; i,
\end{align}
where \(F_i\) may differ across cities and evolve in time. A power-law
trajectory of the form
\begin{equation}
Y_i(t) = \mathcal{C}_i\,P_i(t)^{\beta_i}\qquad \forall t,\; \mathrm{fixed}\; i,
\end{equation}
defines a city-specific exponent \(\beta_i\) and prefactor
\(\mathcal{C}_i\), but this is only a special case. In general, neither the
functions \(F_i\) nor the exponents \(\beta_i\) need to be identical across cities. 

\subsection{From individual cities to urban scaling}

We now ask how the transversal exponent emerges from individual trajectories.
Let \(x_i=\ln P_i(t)\) and \(y_i=\ln Y_i(t)\). At fixed time, ordinary least
squares (OLS) gives
\begin{equation}
    \beta_T =
    \frac{\mathrm{Cov}(x,y)}{\mathrm{Var}(x)} ,
    \label{eq:betaT}
\end{equation}
with \(\mathrm{Cov}(x,y)=N^{-1}\sum_{i=1}^N
(x_i-\langle x\rangle)(y_i-\langle y\rangle)\), and \(\langle\cdot\rangle\) an
average over cities (see also Supplementary section II). This definition is purely cross-sectional and makes no
assumption about the temporal path of any city. Other estimators could be used,
but OLS is common in empirical studies and makes the contributions to
\(\beta_T\) explicit; alternative fits give similar estimates for the cases
studied below, as shown in the SM.

To connect \(\beta_T\) to individual trajectories, we write the local relation
\begin{equation}
    y_i(t) \simeq \alpha_i(t)+\beta_i(t)x_i(t),
    \label{eq:local_longitudinal_relation}
\end{equation}
where \(\beta_i(t)\) is the local longitudinal elasticity. It is estimated by a
rolling regression of \(\ln Y_i\) against \(\ln P_i\) along city \(i\), using a
short window centered on \(t\) -- typically five years in the data below -- and
approximates
\[
    \beta_i(t)=\frac{d\ln Y_i(t)}{d\ln P_i(t)} .
\]
The intercept \(\alpha_i(t)\) gives the local level of the trajectory in the
\((\ln P,\ln Y)\) plane. Substituting Eq.~\eqref{eq:local_longitudinal_relation} into
Eq.~\eqref{eq:betaT} gives
\begin{equation}
    \beta_T
    =
    \langle \beta\rangle
    +
    \frac{\mathrm{Cov}(x,\alpha)}{\mathrm{Var}(x)}
    +
    \frac{\mathrm{Cov}\!\left(x,(\beta-\langle\beta\rangle)x\right)}
         {\mathrm{Var}(x)} ,
    \label{eq:betaT_cov_identity}
\end{equation}
where all quantities are evaluated at the same date \(t\), omitted for
readability (see the SM for details). Thus \(\beta_T\) equals the mean longitudinal elasticity only if the two covariance terms vanish or cancel. Otherwise it is an ensemble-level
quantity shaped by heterogeneity and correlations across cities, not the
dynamical exponent of individual trajectories.

\subsection{Empirical illustrations}

We test this decomposition (Eq.~\ref{eq:betaT_cov_identity}) with three examples. The first is the ``fundamental
allometry'' between urbanized area and population
\cite{ribeiro2023mathematical}, where individual dynamics are approximately
linear or piecewise linear. The second is wages, where trajectories are more
heterogeneous. Congestion delay, previously studied in
\cite{depersin2018global}, is revisited in the SM. For each city we estimate
\(\alpha_i(t)\) and \(\beta_i(t)\), compute the different terms appearing in
Eq.~\eqref{eq:betaT_cov_identity}, and compare the predicted \(\beta_T\) with
the exponent measured directly from cross-sectional fits.

\subsubsection{Area-population relation}

We first consider how built area \(A\) increases with population \(P\), a
classical urban scaling relation discussed in
\cite{rybski2017cities,stewart1947social,nordbeck1971allometric,
woldenberg1973allometric,batty2011defining,ortman2014prehistory,
hamilton2007nonlinear,burger2022global}. Early estimates gave
\(\beta_T\approx 3/4\) \cite{stewart1947social} and
\(\beta_T\approx 2/3\) \cite{nordbeck1971allometric}. Batty and Ferguson
\cite{batty2011defining} reported that most estimates are sublinear (see the SM
for more details), whereas Burger \emph{et al.} \cite{burger2022global} found
exponents near \(5/6\) for half of the countries studied and indistinguishable
from one for the others.

Longitudinal measurements \cite{marquis2025universal} use the Angel
\emph{et al.} dataset for 1800--2000 \cite{angel2012atlas} and the
1985--2015 datasets \cite{wsfevo,worldpop}. In all cases, individual cities are
well described by
\begin{align}
     A_i(t) \approx a_i(t)  P_i(t) \, ,
\end{align}
where \(a_i(t)\) is the inverse density. The constant term is omitted because it
does not affect variations. Trajectories can be strictly linear
(\(a_i=\mathrm{const.}\)) or piecewise linear, with \(a_i=a_1\) before a
city-specific time \(t^*\) and \(a_i=a_2\) after it, usually with \(a_2<a_1\).
Typical linear relations, yearly over 30 years and sporadically over about 200
years, are shown in Fig.~S4 of the SM; density-breaking cases are discussed in
\cite{marquis2025universal}. Power-law fits \(A\sim P^\beta\) are generally
worse than linear or piecewise linear fits and yield dispersed exponents,
\(\beta=1.06\pm0.36\), between \(0.5\) and \(1.8\) (Fig.~S5 in the SM). Thus,
population growth and spatial expansion are approximately linearly coupled
along individual trajectories.

\begin{figure}
  \centering
  \includegraphics[width=0.4\textwidth]{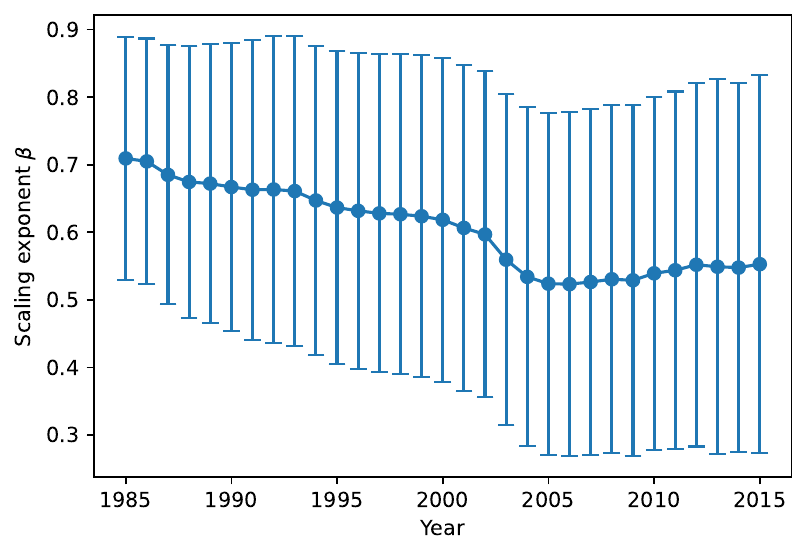}
  \caption{{\bf Cross-sectional scaling: area--population.}
    Temporal evolution of \(\beta_T\) for the 19 cities studied in
    \cite{marquis2025universal}, estimated by OLS in log--log space.
    Alternative estimators (Theil--Sen, weighted least squares) give similar
    results (see Section IV of the Supplementary Material). Error bars indicate the exact $95\%$ confidence interval of the OLS slope estimate, centered around the estimated exponent~$\beta_T$, computed from the 19 cities.}
  \label{fig:areapop}
\end{figure}

In contrast, cross-sectional fits for the same data give exponents different
from one. For the data studied in~\cite{marquis2025universal}, we obtain Fig.~\ref{fig:areapop} showing that 
\(\beta_T\simeq0.5-0.7\), robust to alternative estimators (SM). Read literally,
this would imply that density increases with city size, in contradiction with
the approximately linear longitudinal dynamics. The difference arises because
the inverse density \(a_i\) varies across cities and time, producing covariance
terms in Eq.~\eqref{eq:betaT_cov_identity}. Thus nonlinear cross-sectional scaling
can appear even when each city follows an essentially linear trajectory.

\begin{figure}[htbp!]
    \centering
    \includegraphics[width=1.0\linewidth]{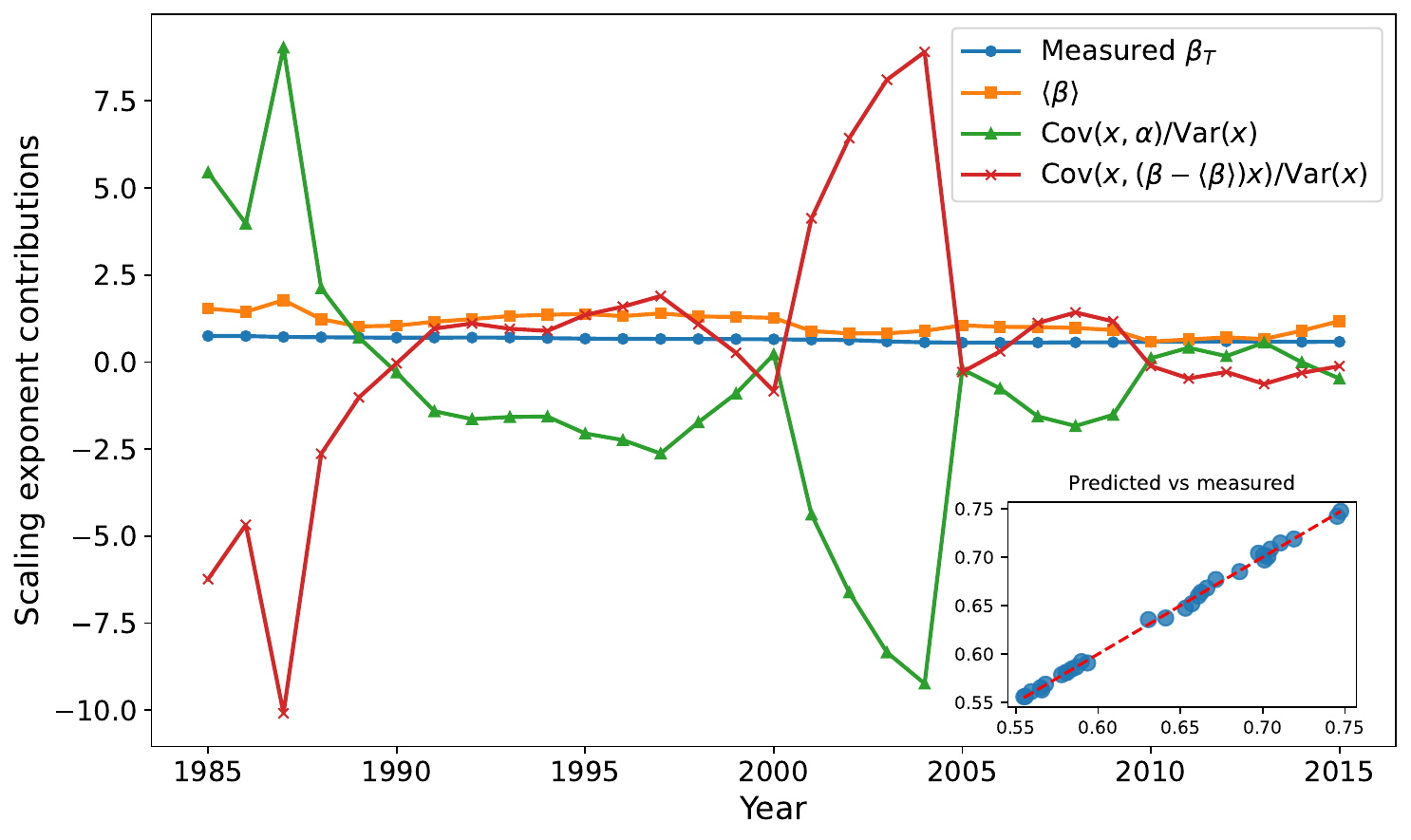}
    \caption{{\bf Statistical contributions to the transversal exponent.} For
    the data discussed in Fig.~\ref{fig:areapop}, the local longitudinal exponent
    \(\beta_i(t)\) is measured for each city. The average \(\langle\beta\rangle\)
    is shown together with the two covariance contributions in
    Eq.~\eqref{eq:betaT_cov_identity}. Their incomplete compensation explains
    the deviation between \(\beta_T\) and \(\langle\beta\rangle\). Inset:
    measured exponent (obtained by directly fitting the area versus population at fixed time) versus the prediction from
    Eq.~\eqref{eq:betaT_cov_identity}.}
    \label{fig:decompostion_avp}
\end{figure}

Figure~\ref{fig:decompostion_avp} confirms this decomposition. The measured
transversal exponent is accurately recovered by summing the mean longitudinal
elasticity and the two covariance terms. The latter have opposite signs and do
not fully cancel, shifting \(\beta_T\) away from the longitudinal value. The cross-sectional exponent thus reflects an aggregate statistical effect over the ensemble of cities rather than a genuine dynamical law governing the growth of individual cities.

To illustrate Eq.~(\ref{eq:betaT_cov_identity}) analytically, we now
consider a particularly simple case in which the longitudinal trajectories
are piecewise linear. This assumption is introduced only to obtain an
analytically tractable example and is not required for the general result,
Eq.~(\ref{eq:betaT_cov_identity}).

Specifically, we show how a cross-sectional exponent different from unity
can emerge even though the longitudinal dynamics of each city are locally
linear. Motivated by the empirically observed piecewise-linear trajectories,
we consider a simple threshold model in which the area $A_i$ of city $i$
depends on its population $P_i$ as
\begin{equation}
    A_i =
    \begin{cases}
        a_1 P_i, & P_i < P_i^*,\\[2mm]
        a_2 P_i + (a_1-a_2)P_i^*, & P_i \geq P_i^* .
    \end{cases}
\end{equation}
The additive term ensures continuity at the city-specific threshold \(P_i^*\), which are drawn from a given distribution. Each city grows linearly in each regime, but a
transversal comparison mixes cities on different branches. Applying
Eq.~\eqref{eq:betaT_cov_identity} to this case when all longitudinal exponents are
\(\beta_i=1\)  (see the SM for details)
\begin{equation}
    \beta_T = 1 + C\,(\ln a_2 - \ln a_1),
\end{equation}
where \(C>0\) depends on the distributions of \(P\) and \(P^*\). Therefore
\(\beta_T>1\) if \(a_2>a_1\), and \(\beta_T<1\) if \(a_2<a_1\). Both the
superlinear exponents observed in Angel \emph{et al.} \cite{angel2012atlas} and
the sublinear exponents in \cite{marquis2025universal} can thus arise from
linear longitudinal dynamics. Simulations of fixed and random thresholds (see the section III of the SM)
confirm that \(\beta_T=1\) is recovered only when cities have identical dynamics
and initial conditions; any source of heterogeneity -- whether in initial populations, in thresholds, or more generally in the dynamical 
parameters -- creates correlations between \(\alpha_i\) and \(x_i\), causing
\(\beta_T\) to depart from the true longitudinal exponent.


\begin{figure*}[t]
    \centering
    \includegraphics[width=0.7\linewidth]{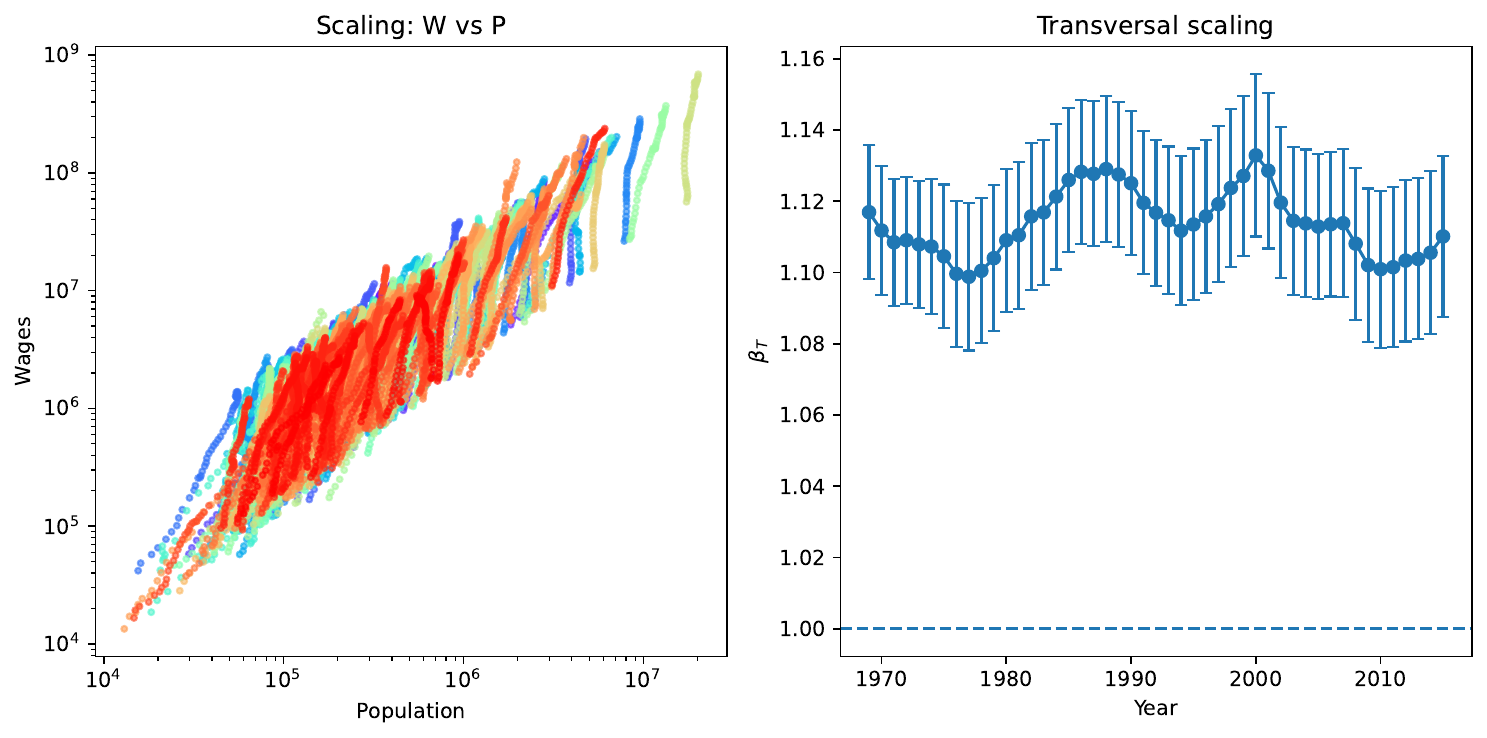}
    \caption{{\bf Wages: transversal scaling.} (Left) Wages as a function of
    population for 363 MSAs observed yearly from 1969 to 2015. Contiguous
    colored points correspond to the same city; the color scale helps distinguish longitudinal growth from cross-sectional variation. 
    (Right) Transversal exponents ($\beta_T$) as a function of time. The exponents were measured using ordinary least-squares in log-log coordinates. The error bars, centered around these points, represent the exact~$95\%$ confidence intervals for the estimated OLS slope, calculated on the~$n=363$ cities. Data from \cite{bettencourt2013origins}.}
    \label{fig:wages}
\end{figure*}

\subsubsection{Wages}

Total wages are a standard example of urban scaling. They have been reported to
grow superlinearly with population \cite{bettencourt2007growth}, often
interpreted as a consequence of enhanced social and economic interactions. In
particular, interaction-based theories predict
\(\beta=7/6\simeq1.17\) \cite{bettencourt2013origins}. For 363 US
Metropolitan Statistical Areas in 2006, the transversal estimate is
\(\beta_T=1.13\pm0.02\), close to this prediction. We use the full dataset of
\cite{bettencourt2013origins}, with yearly wages and population from
1969--2015.

Figure~\ref{fig:wages} shows that the cross-sectional exponent is remarkably
stable over time, as for congestion delay, while individual trajectories are
highly heterogeneous in slope and amplitude (see the distribution of
\(\beta_i\) in the Fig.~S6 in the SM). Decomposing \(\beta_T\) with
Eq.~\eqref{eq:betaT_cov_identity} shows that
\(\langle\beta\rangle\simeq4-5\), far larger than both \(\beta_T\) and the prediction \(7/6\) (section V of the SM).

Individual cities therefore display much steeper wage trajectories than
suggested by the cross-sectional fit. The stability of \(\beta_T\) results from
a balance between heterogeneous longitudinal dynamics and cross-city
correlations, not from a universal trajectory followed by all cities. The
predicted transversal exponent reproduces the measured one (see Fig.~S7 in the SM), with remaining deviations attributable to the large dispersion of individual longitudinal exponents.

More generally, a transversal exponent can always be measured, whatever the
temporal behaviour of individual cities. Some cities may grow rapidly, others
more slowly, and some may not even follow a power law. A cross-sectional fit will
nevertheless return a value of \(\beta_T\). This value measures a statistical
property of the ensemble observed at a given time, not the longitudinal dynamics
of any individual city.

\section{Discussion}

Urban scaling is often used to assess how city size affects quantities such as
CO\(_2\) emissions, crime, patents, infrastructure, or economic output. The
usual approach is to compare many cities at one date and fit how an aggregate
quantity \(Y\) varies with population \(P\). The exponent is then often
interpreted as measuring the effect of population size: if it is larger than
one, the per-capita quantity increases with city size; if it is smaller than
one, the per-capita quantity decreases. This interpretation matters because it
can support broad claims about cities. For instance, sublinear scaling of
CO\(_2\) emissions has sometimes been taken to suggest that larger or denser
cities are intrinsically more environmentally efficient, with possible
implications for planning, density, and governance.

The difficulty is that this interpretation relies on a strong assumption.
Comparing different cities at a single date implicitly treats them as different
stages of the same underlying trajectory. A city with population \(P_1\) and
value \(Y_1\) is then interpreted as a scaled version of another city with
population \(P_2>P_1\) and value \(Y_2\). This would imply that, as the first
city grows from \(P_1\) to \(P_2\), it should approach a value of \(Y\) close to
\(Y_2\). Such a view assumes that cities follow the same dynamical path and
differ mainly by size. This is very restrictive: cities have distinct histories,
spatial constraints, institutions, technologies, and planning trajectories.

Our results show that this assumption is generally not valid. The usual urban
scaling exponent cannot, in general, be interpreted as a direct estimate of how
an individual city changes as it grows. Rather, it is an aggregate quantity
shaped by the distribution of city sizes, the heterogeneity of city-specific
trajectories, local responses at current population levels, and correlations
between these quantities. Cross-sectional scaling therefore reflects not only
city-level dynamics, but also the statistical structure of the urban system as a
whole. The behaviour encoded by the cross-sectional exponent is in general not followed by any individual city.

In some specific cases, such exponents may have a clear interpretation, namely when larger systems are, to a good approximation, scaled-up versions of smaller ones, and when their present structure is not strongly determined by specific
histories. A classical example is Kleiber's
law~\cite{Kleiber1932,Kleiber1947,Enquist1998-py,West1997}, which relates
metabolic rate to body mass through \(B\sim M^{3/4}\). In that case, the
exponent can be interpreted because organisms of different sizes share
comparable biological constraints, in particular constraints associated with
transporting energy and resources through the body. 

For cities, the situation is different. Scaling exponents are sometimes used to
support claims that larger cities are more innovative, more efficient, or
greener. Our results show that such conclusions should be drawn with caution. A
sublinear or superlinear transversal exponent does not imply that increasing the
population of a given city will produce the corresponding decrease or increase
in the quantity per capita. It may instead reflect differences between cities:
their histories, densities, infrastructures, economic structures, planning
regimes, or geographic constraints. Using cross-sectional scaling laws for
prediction is especially problematic. For example, Ref.~\cite{HANKEY20104880}
combines exogenous population-growth scenarios with cross-sectional scaling
relations to forecast future greenhouse-gas emissions and travelled distances
under different sprawling scenarios. Such approaches implicitly assume that
scaling laws observed across cities remain valid along the future trajectories
of individual cities, which can lead to misleading predictions and policy
conclusions.

From a policy perspective, this calls for a shift in how urban scaling laws are interpreted. Cross-sectional scaling laws should not be treated as fundamental laws of city growth, nor used alone as guides for urban design or governance. They can reveal statistical regularities at the level of a system of cities, but they do not by themselves prescribe how a particular city will evolve, or how it should be planned. Cross-sectional exponents are derived quantities: they emerge from the ensemble of longitudinal trajectories followed by individual cities. The fundamental objects are therefore these trajectories themselves---how emissions, infrastructure, land use, innovation, congestion, or housing costs change as a given city grows. Understanding urban growth requires starting from these temporal trajectories, because they contain the information needed to assess the effects of planning decisions, technological change, institutional constraints, and path dependence.

\section*{Data availability}

All data used in this paper are publicly and freely available from the sources
cited in the manuscript. The area--population data are from the datasets
described in~\cite{angel2012atlas,wsfevo,worldpop},
and the wage data are from Ref.~\cite{bettencourt2013origins}. No restricted or proprietary data were used.

\section*{Acknowledgements}

MB thanks the Complexity Science Hub in Vienna and R. Prieto-Curiel, as well as the University of Padua and Manlio De Domenico, for their invitations, for the opportunity to present and discuss this work, and for many interesting and stimulating discussions.

\section*{Funding Statement}

The authors received no specific funding for this work.

\section*{Author Contributions Statement}

UM conceived the study, developed the analysis code and theoretical framework, and wrote the manuscript. MB conceived the study, developed the theoretical framework, performed the analysis, and wrote the manuscript.

\section*{Competing Interests Statement}

The authors declare no competing interests.


\bibliography{final/ref_scaling}
\bibliographystyle{IEEEtran}

\end{document}






\maketitle

\tableofcontents


\section{Earlier approaches: Temporal versus transversal scaling}

The relation between transversal (cross-sectional) and longitudinal (or temporal) scaling has been examined in several recent works \cite{keuschnigg2019scaling,bettencourt2020interpretation,lobo2025urban,ribeiro2020relation,connor2024frontier,connor2025big}. These studies consistently show that the exponents inferred from the temporal evolution of individual cities can differ substantially from the cross-sectional exponents obtained by comparing cities at a given time \cite{keuschnigg2019scaling,connor2025big}. Yet they generally continue to treat the transversal exponent as the main quantity of interest for characterizing urban scaling.

In particular, in~\cite{bettencourt2020interpretation}, the authors establish a mapping between temporal and cross-sectional scaling forms. While mathematically correct, this construction raises several conceptual and interpretational issues. The starting point of temporal scaling discussed in this paper is to assume that the individual city behavior is described by
\begin{equation}
    Y_i(t) = a_i P_i(t)^{\beta_i} e^{\chi_i(t)},
\label{eq:assum}
\end{equation}
with a time-independent, city-specific prefactor $a_i$ (the residual $\chi_i$ for the temporal fit vanishes for each city under time averaging). This assumption is not innocuous: it effectively absorbs all temporal variations unrelated to population growth into the longitudinal exponent $\beta_i$. As a consequence, the exponent~$\beta_i$ is not a pure measure of scale effects, but a composite quantity mixing scale dependence, temporal trends, and fluctuations. This point is made explicit in ~\cite{bettencourt2020interpretation}, where $\beta_i$ is shown to depend on the temporal evolution of the prefactor, on variations of the cross-sectional exponent $\beta_T(t)$, and on residual terms. More precisely, using this assumption (Eq.~\eqref{eq:assum}), it can be shown that the temporal scaling is fundamentally ill-conditioned when population growth is small. In particular, in the simplified case of constant growth rates (a similar relation is also derived in~\cite{ribeiro2020relation}),
\begin{equation}
    \beta_i \simeq \beta_T + \frac{\eta}{\gamma_i},
    \label{eq:bettencourt}
\end{equation}
so that $\beta_i$ diverges as $\gamma_i \to 0$ (here, $\eta$ and $\gamma_i$ denote the assumed constant growth rates of the prefactor and population of city $i$, respectively). This divergence is not a property of the underlying urban dynamics, but an artefact connected to the specific form chosen for the temporal scaling (Eq.~\eqref{eq:assum}). As emphasized in \cite{bettencourt2020interpretation}, this leads to unstable and even arbitrarily large exponents for slowly growing or shrinking cities, a phenomenon absent in the examples that we will discuss below. 

Although the authors present the mapping between temporal and cross-sectional exponents as an equivalence, this equivalence is purely formal. The conditions under which transversal and longitudinal exponents coincide are extremely restrictive: absence of intensive growth, time-independent cross-sectional exponents, and negligible residuals. These conditions are rarely satisfied in empirical urban systems, where intensive growth and heterogeneity are ubiquitous. In practice, the temporal exponent therefore does not measure the same quantity as the transversal exponent, despite the formal mapping. 

As noted above, this discussion relies crucially on the assumption that the evolution of $Y_i(t)$ involves a city-specific prefactor $a_i$ that is independent of time. Empirical evidence suggests that this assumption does not always hold. In particular, as we will discuss below, many cities follow well-defined trajectories in the area ($A$) versus population ($P$) plane of the form \cite{marquis2025universal}
\begin{equation}
A_i(t) \simeq a_i(t) P_i(t) \,.
\end{equation}
In some cases, $a_i$ is approximately constant in time, while in others it exhibits a piecewise behavior, taking values $a_1$ for $t < t^*$ and $a_2$ for $t > t^*$, where $t^*$ is city-dependent. Such piecewise linear relations are naturally associated with different phases of urban development. Under these conditions, the divergence mechanism discussed in~\cite{bettencourt2020interpretation} does not arise. The temporal evolution remains effectively linear, with exponent $\beta_i = 1$, while the dynamics are captured by a time-dependent prefactor.

More fundamentally, the framework proposed in~\cite{bettencourt2020interpretation} takes the cross-sectional scaling relation as the primary object and infers temporal behavior from it. In this view, the transversal scaling law is treated as fundamental, while temporal exponents are interpreted as derived quantities. Here, we propose the opposite perspective: the dynamics of individual cities are taken as fundamental, and transversal scaling is viewed as an emergent statistical pattern arising from an ensemble of heterogeneous trajectories. From this standpoint, the observed cross-sectional exponent does not encode a universal growth law, but reflects a combination of intrinsic city dynamics, inter-city heterogeneity, and correlations within the urban system.

In this context, the key issue is not the instability of temporal scaling, but the diversity of city dynamics. Different cities follow distinct relations $Y_i(P)$ shaped by their history, spatial constraints, and institutional context. The exponent measured from cross-sectional data is therefore an aggregate statistical quantity rather than a dynamical invariant. Understanding how transversal scaling emerges from this heterogeneity is essential for a correct interpretation of urban scaling laws.

 Moreover, in~\cite{ribeiro2020relation}, the authors analyze both transversal 
 \begin{align} \label{eq:beta0}
     Y_i(t) = Y_0(t) P_i(t)^\beta
 \end{align} and longitudinal scaling across more than $5000$ Brazilian cities and observe clear discrepancies between the corresponding exponents. They argue, however, that these differences can be reconciled by removing so-called “external factors,” namely the temporal variation of the cross-sectional prefactor $Y_0(t)$. Once this correction is applied, the longitudinal exponents appear, in some cases, to cluster around the transversal one. As noted by the authors, however, there is no general agreement between the distribution of the decomposed exponents and the value of the transversal exponent; this correspondence appears to hold for GDP, but not for the water network length \cite{ribeiro2020relation}. To address this issue, they propose an alternative method, similar to that of \cite{bettencourt2020interpretation}, and obtain a relation between longitudinal and transversal exponents akin to Eq.~\eqref{eq:bettencourt}.

 In ~\cite{hong2020} Hong et al. investigate employment trajectories across cities and over time, and argue for the existence of a universal pathway of urban economic development. According to their findings, cities transition from economies dominated by manual labor to ones centered on the tertiary sector, with an estimated threshold of approximately $1.2 \cdot 10^6$ inhabitants in the United States. They introduce the concept of urban recapitulation, whereby observing a single city over time yields patterns equivalent to comparing different cities of different sizes at a fixed point in time. To support this claim, they define a recapitulation score
\begin{align}
S = 1 - \frac{|\hat{\beta} - \beta|}{\beta},
\end{align}
where $\hat{\beta}$ is a longitudinal exponent obtained after removing the nation-wide growth factor, i.e., the transversal scaling prefactor $Y_0(t)$ ($\beta$ is the transversal exponent of Eq.~\ref{eq:beta0}). By construction, $S$ is large when transversal and longitudinal exponents are close. They report an average value of about $0.7$, which they interpret as strong evidence of agreement between the two. We argue that this interpretation is biased. The apparent agreement between exponents only emerges after normalizing by a system-wide quantity, namely $Y_0(t)$, and is therefore not intrinsic to the dynamics of individual cities. More broadly, we show in the following that the transversal scaling hypothesis leads to a distorted view of urban growth: transversal exponents do not possess a clear longitudinal dynamical meaning and may, in fact, be misleading when used to infer or predict the evolution of individual cities.

\section{Methods}

\subsection{OLS estimate of the transversal exponent}

At a given date, we write the transversal relation in log--log form as
\begin{equation}
y_i = \alpha_T + \beta_T x_i + \varepsilon_i,
\end{equation}
with $x_i=\ln P_i$ and $y_i=\ln Y_i$.
The OLS estimate minimizes the total deviation given by
\begin{equation}
S(\alpha_T,\beta_T)=\sum_{i=1}^N (y_i-\alpha_T-\beta_T x_i)^2 .
\end{equation}
The minimization equations $\partial S/\partial \alpha_T=\partial S/\partial \beta_T=0$ imply
\begin{equation}
\alpha_T=\bar y-\beta_T \bar x
\end{equation}
and
\begin{equation}
\sum_{i=1}^N (x_i-\bar x)(y_i-\bar y)
=
\beta_T \sum_{i=1}^N (x_i-\bar x)^2 .
\end{equation}
Hence
\begin{equation} \label{eq:betaT}
\beta_T
=
\frac{\sum_i (x_i-\bar x)(y_i-\bar y)}
{\sum_i (x_i-\bar x)^2}
=
\frac{\mathrm{Cov}(x,y)}{\mathrm{Var}(x)} .
\end{equation}
Thus, the transversal exponent estimated by OLS is the ratio between the cross-sectional covariance of \(x=\ln P\) and \(y=\ln Y\), and the cross-sectional variance of \(x\).

\subsection{Goodness of fit}

In addition to the expression of the transversal exponent, the quality of the fit can also be written in a simple form. For the OLS regression
\begin{equation}
y_i = \alpha + \beta x_i + \varepsilon_i,
\end{equation}
the coefficient of determination is given by
\begin{equation}
R^2
=
\frac{\mathrm{Cov}(x,y)^2}{\mathrm{Var}(x)\,\mathrm{Var}(y)}
=
\rho_{xy}^2,
\end{equation}
where \(\rho_{xy}\) is the Pearson correlation coefficient computed over the cross-section. Using the expression \(\beta_T = \mathrm{Cov}(x,y)/\mathrm{Var}(x)\), this can also be written as
\begin{equation}
R^2 = \beta_T^2 \,\frac{\mathrm{Var}(x)}{\mathrm{Var}(y)}.
\end{equation}
This relation shows that, while the exponent \(\beta\) is determined by the covariance between \(x\) and \(y\), the goodness of fit depends on how this covariance compares to the total variance of \(y\). In particular, a well-defined exponent can coexist with a low value of \(R^2\) if the variance of \(y\) is large. In the context of transversal scaling, this highlights that the existence of a scaling exponent does not necessarily imply a tight functional relationship, but rather reflects the underlying covariance structure of the ensemble of cities.

\subsection{Transversal exponent calculations}
 
\subsubsection{Individual power-law case}

We start with the simplest case (although probably not the most common) where individual cities behave as power laws with various exponents
\begin{align}
    Y_i(t)=a_iP_i(t)^{\beta_i}
\end{align}
where the prefactor $a_i$ could depend on time or population, and the exponent $\beta_i$ varies from a city to another. In this case, we obtain
\begin{align}
    y_i(t)=\alpha_i+\beta_i x_i(t)
\end{align}
where $\alpha_i=\ln a_i$. Using the estimate (eq.~\eqref{eq:betaT}) of $\beta_T$, we find
\begin{equation} \label{eq:indiv}
\beta_T(t)=
\frac{\mathrm{Cov}(x,\alpha)}{\mathrm{Var}(x)}
+
\frac{\mathrm{Cov}(x,\beta x)}{\mathrm{Var}(x)}.
\end{equation}

The first term in Eq.~\ref{eq:indiv} corresponds to the OLS slope obtained when regressing the prefactor~$\alpha$ against~$x$. The second term is more subtle. In the special case where all cities share the same exponent, $\beta_i = \beta$, it reduces simply to~$\beta$. In that situation, the difference between the longitudinal exponent~$\beta$ and the transversal exponent~$\beta_T$ is entirely captured by the first covariance term. This shows that $\beta_T$ does not purely reflect intrinsic city dynamics, but is also influenced by correlations between prefactors and population. In the absence of such correlations, one recovers $\beta_T = \beta$. When the exponents $\beta_i$ are heterogeneous, however, the situation becomes more intricate.

Altogether, the transversal exponent emerges as a combination of the average longitudinal exponent and additional statistical contributions that depend on the joint distribution of $\beta$ and $x$. As a result, $\beta_T$ is generally a nontrivial aggregate quantity, whose interpretation is not straightforward.

\subsubsection{General case} \label{subsec:general}

To connect in the general case the cross-sectional scaling exponent to the longitudinal evolution of individual cities, we estimate for each city $i$ and year $t$ a local longitudinal relation in logarithmic variables,
\begin{align}
y_i(t) &\approx \alpha_i(t)+\beta_i(t)x_i(t),
\label{eq:local_longitudinal_relation}
\end{align}
Here $\beta_i(t)$ is the local longitudinal elasticity of city $i$, estimated from a rolling regression of $\ln Y_i$ against $\ln P_i$ along the temporal trajectory of that city, while $\alpha_i(t)$ is the corresponding local intercept. More precisely, for each city and each year $t$, we consider a short time window centered on $t$ (typically five years for the data studied below) and perform a linear regression of $\ln Y_i$ on $\ln P_i$ using the observations within this window. The resulting slope provides a local estimate of the derivative
\begin{align}
    \beta_i(t)=\frac{d\ln Y_i}{d\ln P_i}
\end{align}
which defines the instantaneous longitudinal elasticity $\beta_i(t)$, while the intercept defines \(\alpha_i(t)\). This procedure provides a local linear approximation of the city trajectory in the $(\ln P,\ln Y)$ plane, allowing us to capture gradual changes in scaling behavior over time while reducing the noise that would arise from estimating the derivative from only two consecutive observations.

At a fixed time \(t\), we define the average over the \(N\) cities observed at that time as
\[
    \langle z\rangle=\frac{1}{N}\sum_{i=1}^N z_i .
\]
The covariance and variance are therefore
\[
    \mathrm{Cov}(x,y)
    =
    \left\langle
    \left(x-\langle x\rangle\right)
    \left(y-\langle y\rangle\right)
    \right\rangle ,
    \qquad
    \mathrm{Var}(x)
    =
    \left\langle
    \left(x-\langle x\rangle\right)^2
    \right\rangle .
\]
The transversal exponent is the ordinary least-squares slope of the
cross-sectional relation between \(y_i\) and \(x_i\), hence
\begin{equation}
    \beta_T(t)
    =
    \frac{\mathrm{Cov}(x,y)}{\mathrm{Var}(x)} .
    \label{eq:si_betaT_ols}
\end{equation}

We now substitute the local longitudinal relation of each city Eq.~\ref{eq:local_longitudinal_relation} into Eq.~\eqref{eq:si_betaT_ols} and obtain
\begin{align}
    \beta_T(t)
    &=
    \frac{\mathrm{Cov}\!\left(x,\alpha+\beta x\right)}
         {\mathrm{Var}(x)} .
\end{align}
Using the linearity of the covariance in its second argument, we obtain
\begin{align}
    \beta_T(t)
    &=
    \frac{\mathrm{Cov}(x,\alpha)}
         {\mathrm{Var}(x)}
    +
    \frac{\mathrm{Cov}(x,\beta x)}
         {\mathrm{Var}(x)} .
    \label{eq:si_betaT_first_decomposition}
\end{align}

We then decompose the local elasticities into their average and their
city-specific deviations:
\[
    \beta_i
    =
    \langle \beta\rangle
    +
    \left(\beta_i-\langle\beta\rangle\right).
\]
Thus,
\[
    \beta_i x_i
    =
    \langle\beta\rangle x_i
    +
    \left(\beta_i-\langle\beta\rangle\right)x_i .
\]
Substituting this expression into the second covariance term gives
\begin{align}
\nonumber
    \mathrm{Cov}(x,\beta x)
    &=
    \mathrm{Cov}
    \left(
        x,
        \langle\beta\rangle x
        +
        \left(\beta-\langle\beta\rangle\right)x
    \right) \\
    &=
    \mathrm{Cov}\left(x,\langle\beta\rangle x\right)
    +
    \mathrm{Cov}
    \left(
        x,
        \left(\beta-\langle\beta\rangle\right)x
    \right).
\end{align}
Since \(\langle\beta\rangle\) is a constant across cities at fixed time \(t\),
\begin{align}
    \mathrm{Cov}\left(x,\langle\beta\rangle x\right)
    =
    \langle\beta\rangle \mathrm{Cov}(x,x)
    =
    \langle\beta\rangle \mathrm{Var}(x) \,.
\end{align}
Therefore,
\begin{align}
    \mathrm{Cov}(x,\beta x)
    &=
    \langle\beta\rangle \mathrm{Var}(x)
    +
    \mathrm{Cov}
    \left(
        x,
        \left(\beta-\langle\beta\rangle\right)x
    \right).
\end{align}

Inserting this result into Eq.~\eqref{eq:si_betaT_first_decomposition}, we find
\begin{align}
    \beta_T(t)
    &=
    \frac{\mathrm{Cov}(x,\alpha)}
         {\mathrm{Var}(x)}
    +
    \frac{
    \langle\beta\rangle \mathrm{Var}(x)
    +
    \mathrm{Cov}
    \left(
        x,
        \left(\beta-\langle\beta\rangle\right)x
    \right)}
    {\mathrm{Var}(x)} \\
    &=
    \langle\beta\rangle
    +
    \frac{\mathrm{Cov}(x,\alpha)}
         {\mathrm{Var}(x)}
    +
    \frac{
    \mathrm{Cov}
    \left(
        x,
        \left(\beta-\langle\beta\rangle\right)x
    \right)}
    {\mathrm{Var}(x)} .
\end{align}
We thus obtain
\begin{equation}
    \boxed{
    \beta_T(t)
    =
    \langle\beta(t)\rangle
    +
    \frac{\mathrm{Cov}(x(t),\alpha(t))}
         {\mathrm{Var}(x(t))}
    +
    \frac{
    \mathrm{Cov}
    \left(
        x(t),
        \left[\beta(t)-\langle\beta(t)\rangle\right]x(t)
    \right)}
    {\mathrm{Var}(x(t))}
    } .
    \label{eq:betaT_cov_identity}
\end{equation}

This expression shows that the transversal exponent can be reconstructed from
the longitudinal behaviour of individual cities, but that it is not, in general,
equal to the mean longitudinal elasticity $\langle\beta(t)\rangle$. Besides the average two correction terms enter: one involving the covariance between city size and
the local intercepts \(\alpha_i(t)\), and one involving the covariance between
city size and deviations of the local elasticities from their mean. These terms
arise from heterogeneity across cities and from the time dependence of the local
parameters \(\alpha_i(t)\) and \(\beta_i(t)\).

Thus, \(\beta_T(t)=\langle\beta(t)\rangle\) only when these covariance terms
vanish or cancel. A clean transversal power law can therefore be given a direct
dynamical interpretation only under restrictive conditions: individual cities
must follow sufficiently homogeneous trajectories, and correlations between city
size and dynamical parameters must remain weak. In most empirical situations,
these conditions are unlikely to be satisfied. Transversal exponents should
therefore be interpreted with caution: rather than revealing universal laws of
urban growth, they often capture effective ensemble-level properties of a
heterogeneous system of cities.

\subsubsection{Two slopes case}
 
We consider here the case when the individual cities have a piecewise linear behaviour with slope $a_1$ for $x<x^*$ and slope $a_2$ for $x>x^*$.
We will compute the transversal exponent using the general OLS equation discussed in the text
\begin{align}
\beta_T(t)=\frac{\mathrm{Cov}(x,\alpha)}{\mathrm{Var}(x)}
+\frac{\mathrm{Cov}(x,\beta x)}{\mathrm{Var}(x)}
\end{align}
We consider that the logarithm $x$ of the population is distributed according to distribution $P(x)$. The area can thus be written in this case as 
\begin{align}
A=
\begin{cases}
  a_1 P & P<P^*\\
  a_2 P + b & P>P^*
\end{cases}
\end{align}
where $b=(a_1-a_2)P^*$ ensures the continuity of $A$. We will assume that we are interested in regimes where $P$ is large enough so that we can neglect $b$. In all cases we have an elasticity (local slope) equal to $1$, possibly with $1/P$ corrections. We write $\alpha_i=\ln a_i$ and we can write for city $i$
\begin{align}
    \alpha_i\approx \alpha_1+(\alpha_2-\alpha_1)\,\theta(x_i-x_i^*)
\end{align}
(neglecting $1/P$ terms) where $x_i^*$ corresponds to $P_i^*$, the slope-changing population for city $i$ ($\theta(x)$ is the Heaviside function). We assume that this threshold is distributed according to $\rho(x^*)$, independently of the current population $x$.
 
The elasticity $\beta_i=1$ (with possible $1/P$ corrections) and we therefore have $\mathrm{Cov}(x,\beta x)/\mathrm{Var}(x)=1$. We therefore get
\begin{align}
\beta_T(t)=1+\frac{\mathrm{Cov}(x,\alpha)}{\mathrm{Var}(x)}
\end{align}
The covariance can be written as
\begin{align}
    \mathrm{Cov}(x,\alpha)&=\langle x\alpha\rangle-\langle x\rangle \langle \alpha\rangle\nonumber\\
    &=(\alpha_2-\alpha_1)\int dx^*\,\rho(x^*)\int_{x^*}^{\infty}dx\,P(x)\,(x-\langle x\rangle)
\end{align}
We denote by
\begin{align}
\mathcal{I}=\int dx^*\,\rho(x^*)\int_{x^*}^{\infty}dx\,P(x)\,(x-\langle x\rangle)
\end{align}
and it is easy to show that $\mathcal{I}\geq 0$. Indeed, the inner integral $\int_{x^*}^{\infty}dx\,P(x)\,(x-\langle x\rangle)$ is non-negative for all $x^*>0$: since $\int_0^\infty dx\,P(x)\,(x-\langle x\rangle)=0$ by definition of the mean, it equals $-\int_0^{x^*}dx\,P(x)\,(x-\langle x\rangle)$, which is non-negative whether $x^*$ is below or above $\langle x\rangle$. Since $\rho(x^*)\geq 0$, the outer integral $\mathcal{I}$ is non-negative as well.
 
The equation for $\beta_T$ thus gives us
\begin{align}
    \beta_T=1+C(\alpha_2-\alpha_1)
\end{align}
where $C=\mathcal{I}/\mathrm{Var}(x)>0$. We thus obtain the result discussed in the text: when $a_2>a_1$ (i.e.\ $\alpha_2>\alpha_1$) we have $\beta_T>1$---a superlinear behaviour, and conversely.

\section{Simulation: the role of dynamics heterogeneity}
\label{sec:heterogeneity}

We perform here a simple simulation in order to show the importance of heterogeneity in the dynamics of cities and how it impacts the value of the transversal exponent, even when the local dynamics are described by the same exponent (here $\beta_i=1$).

We consider $N$ cities with populations $P_i(t) = P_i^{(0)} + t$, where the initial populations $P_i^{(0)}$ are drawn uniformly at random in the interval $[P_{\min}^{(0)},\, P_{\max}^{(0)}]$. The area of city $i$ is given by $A_i(t) = a_i(t)\,P_i(t)$, where the prefactor switches at a threshold: $a_i = a_1$ if $P_i < P_i^*$ and $a_i = a_2$ otherwise (we also impose the continuity of the area at $P=P^*$ which gives an affine function in the second regime). The local elasticity is $\beta_i = 1$ for every city at all times. A cross-sectional OLS regression of $\ln A_i$ on $x_i = \ln P_i$ at fixed $t$ yields the transversal exponent
\begin{equation}
  \beta_T(t) = 1 + \frac{\mathrm{Cov}(\alpha,\, x)}{\mathrm{Var}(x)}\,,
  \label{eq:betaT_2}
\end{equation}
where $\alpha_i = \ln a_i(t)$. This expression shows that any deviation from the true elasticity is controlled entirely by the cross-sectional correlation between log-prefactors and log-populations.

\begin{figure}
  \centering
  \includegraphics[width=\textwidth]{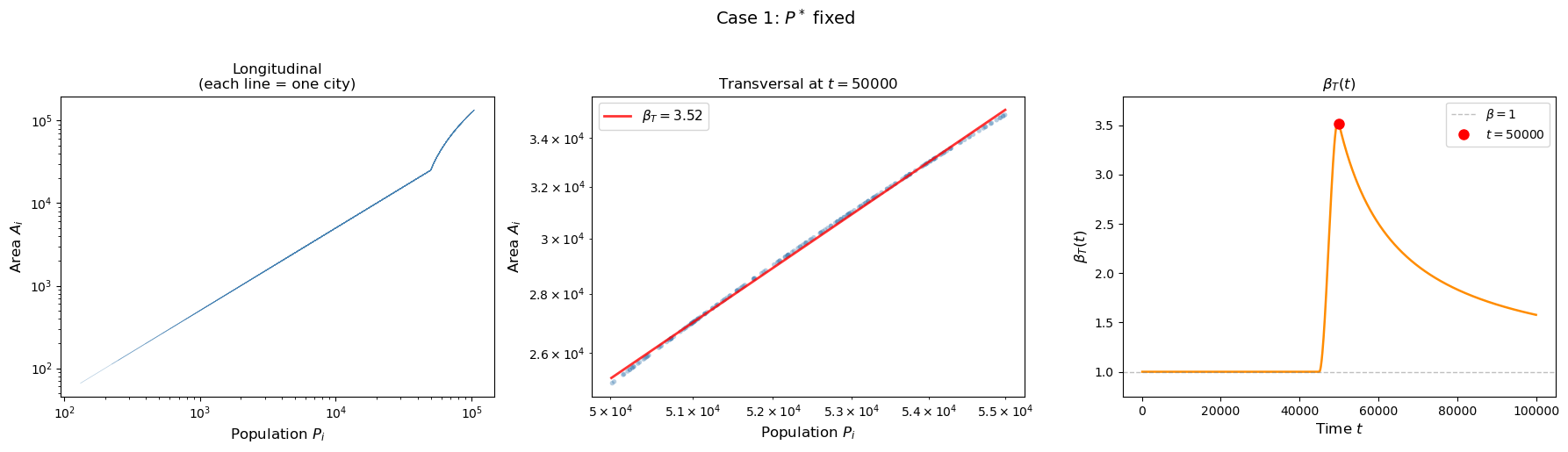}\\[4pt]
  \includegraphics[width=\textwidth]{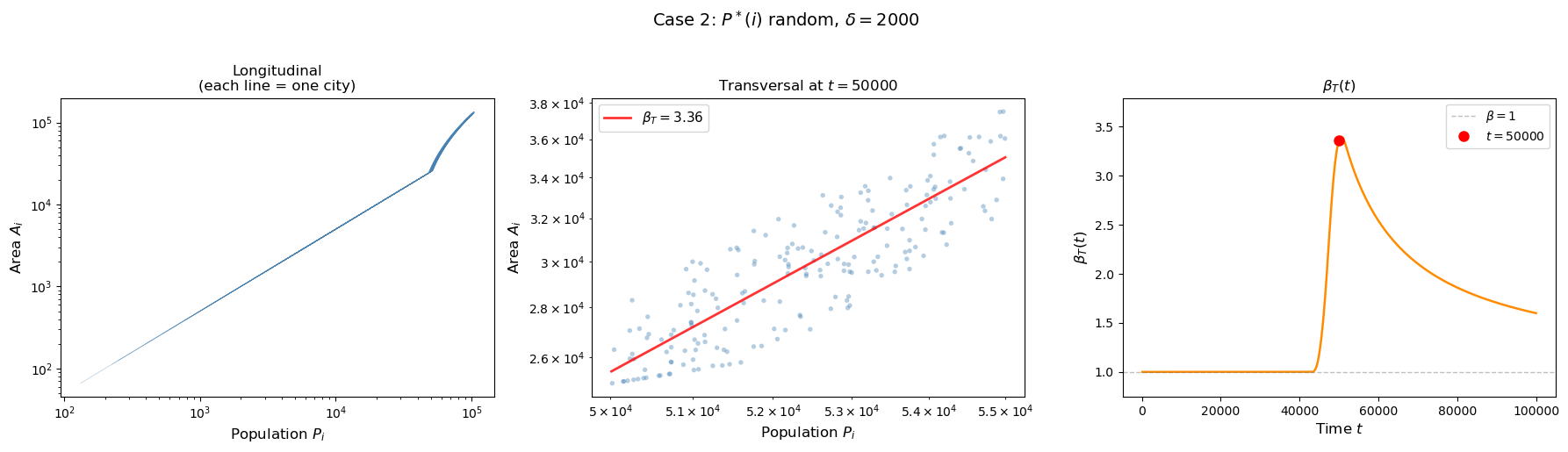}
  \caption{Left: longitudinal trajectories of individual cities 
  ($A_i$ vs.\ $P_i$ as $t$ increases). Middle: transversal snapshot 
  at $t = 50{,}000$ with OLS fit (red line). Right: $\beta_T(t)$; the 
  red dot marks the snapshot time. Top: homogeneous case ($P_i^* = P_{\max}/2$ 
  for all cities); $\beta_T \simeq 1$ except during a narrow transition. 
  Bottom: heterogeneous case ($P_i^*$ random with half-width $\delta = 2000$); 
  $\beta_T$ departs broadly from the true elasticity $\beta = 1$.}
  \label{fig:scaling}
\end{figure}

When all cities share a common threshold $P_i^* = P^*$, then at most times $t$ either all cities satisfy $P_i(t) < P^*$ or all satisfy $P_i(t) \geq P^*$; the prefactor is uniform across cities, $\mathrm{Cov}(\alpha, x) = 0$, and $\beta_T = 1$ exactly (Fig.~\ref{fig:scaling}, top row). However, because the initial populations $P_i^{(0)}$ are heterogeneous (spread over the interval $[P_{\min}^{(0)},\, P_{\max}^{(0)}]$), cities do not all cross $P^*$ at the same time: larger cities reach the threshold first. During the narrow transition window of width $\Delta t = P_{\max}^{(0)} - P_{\min}^{(0)}$, $\alpha_i$ is a step function of $x_i$ and $\beta_T$ spikes sharply. If all cities started with the same initial population $P_i^{(0)} = P_0$, they would cross $P^*$ simultaneously and $\beta_T$ would remain equal to~$1$ at all times. Even in this simple case, heterogeneity---here in initial conditions---is responsible for the departure from the true exponent.

We obtain similar results when each city draws its own threshold $P_i^* \sim \mathcal{U}(P^* - \delta,\, P^* + \delta)$. Cities with larger $P_i$ are more likely to have crossed their individual $P_i^*$, inducing a positive $\mathrm{Cov}(\alpha, x)$ and pushing $\beta_T$ above~$1$ during the transition period (Fig.~\ref{fig:scaling}, bottom row).

The message resulting from this simple simulation is clear: $\beta_T$ recovers the true elasticity $\beta_i=1$ only when all cities follow identical dynamics \emph{and} share the same initial conditions. Any source of heterogeneity---whether in the initial populations, in the transition thresholds, or more generally in the dynamical parameters---can induce cross-sectional correlations between $\alpha_i$ and $x_i$, causing $\beta_T$ to depart from the (true) longitudinal exponent. The decomposition~\eqref{eq:betaT_2} serves as a diagnostic: a nonzero covariance term signals that the measured transversal exponent is contaminated by heterogeneous dynamics rather than reflecting a genuine scaling law.


\section{Changing fitting procedure}

\begin{figure}
    \centering
    \includegraphics[width=0.49\linewidth]{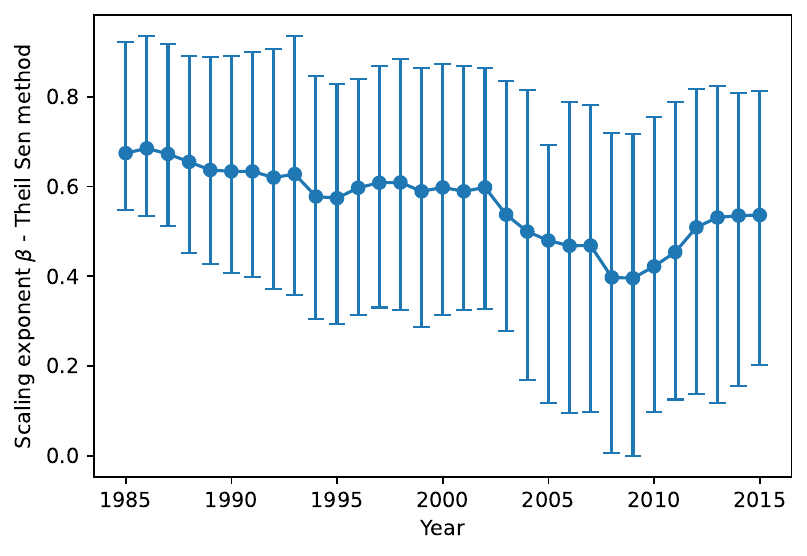}
    \includegraphics[width=0.49\linewidth]{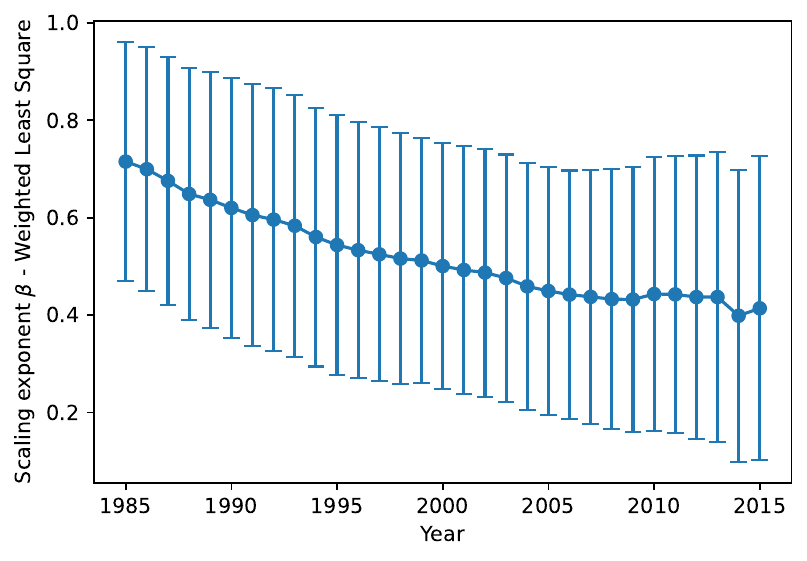}
    \caption{Comparison of transversal exponent estimates using two robust fitting methods. Left panel: the Theil–Sen regression computes the slope of a dataset as the median of all pairwise slopes between points, which reduces its sensitivity to outliers. Right panel: Weighted Least Squares (WLS), which assigns weights to account for heteroskedasticity, giving more reliable estimates when variance varies across data points. }
    \label{fig:fitting}
\end{figure}

While the Ordinary Least Squares (OLS) method in log-log space provides an analytical expression for the transversal exponent, it also has certain limitations: it is sensitive to outliers and relies on the assumption of homoskedasticity. Alternative approaches are more robust against these issues. For instance, the Theil–Sen regression is a linear model that is resilient to extreme values, whereas weighted least squares (WLS) account for heteroskedasticity by assigning different weights to data points. Using the area–population data as an example, Fig.~\ref{fig:fitting} illustrates the exponents and associated uncertainty ranges estimated with these methods.

While the exact exponents value differ, the results are qualitatively similar to those found using the OLS -- both in terms of exponent values and confidence interval widths.

\section{Empirical results}

\subsection{The fundamental allometry}

\begin{figure}
    \centering
    \includegraphics[width=0.7\linewidth]{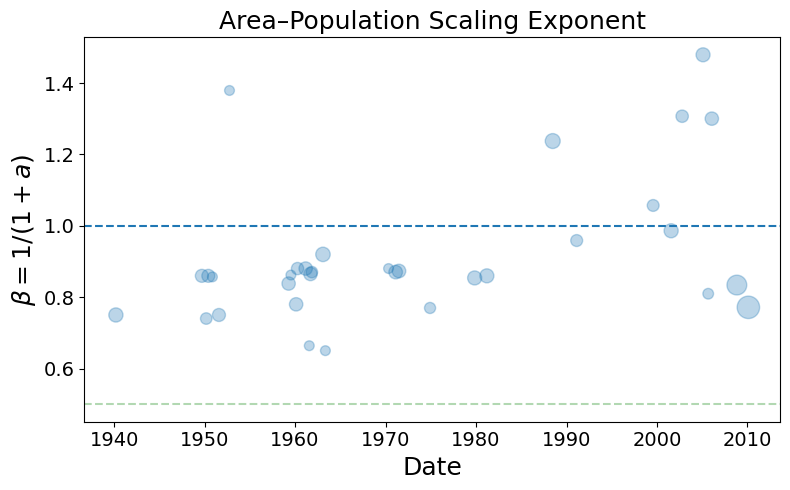}
    \caption{Compilation of transversal exponents $\beta_T$ for the area--population relation across different datasets (from \cite{batty2011defining}). The horizontal axis indicates the reference date of each dataset (the size of symbols refer to the number of cities in the dataset).}
    \label{fig:batty}
\end{figure}

In \cite{batty2011defining}, Batty and Ferguson compile estimates of the transversal exponent $\beta_T$ for the area--population relation reported across a range of studies. These results are summarized in Fig.~\ref{fig:batty}, where the horizontal axis indicates the reference date of each dataset (i.e., the time at which the cross-sectional measurements are performed).

Several features are immediately apparent. First, the reported values of $\beta_T$ display substantial variability across datasets. While most estimates satisfy $\beta_T < 1$, indicating sublinear scaling, a significant number of cases exhibit $\beta_T > 1$. Overall, no clear universal value or systematic trend emerges from these measurements. 

These results do not point to a universal exponent, but rather highlight the strong variability of transversal estimates, consistent with their sensitivity to heterogeneity and correlations across cities.

\subsubsection{Area-population curves: linear vs power-law fit}

\begin{figure}
    \centering
    \includegraphics[width=0.49\linewidth]{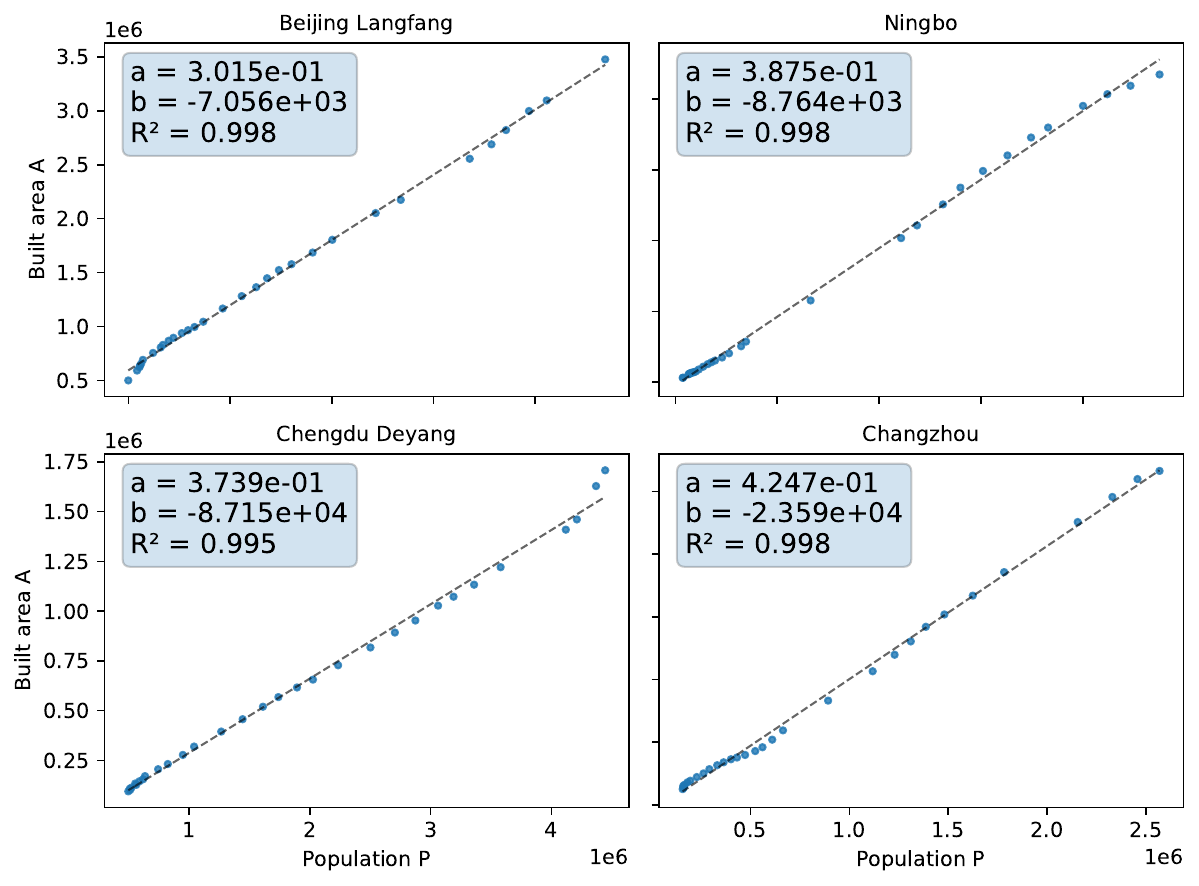}
    \includegraphics[width=0.49\linewidth]{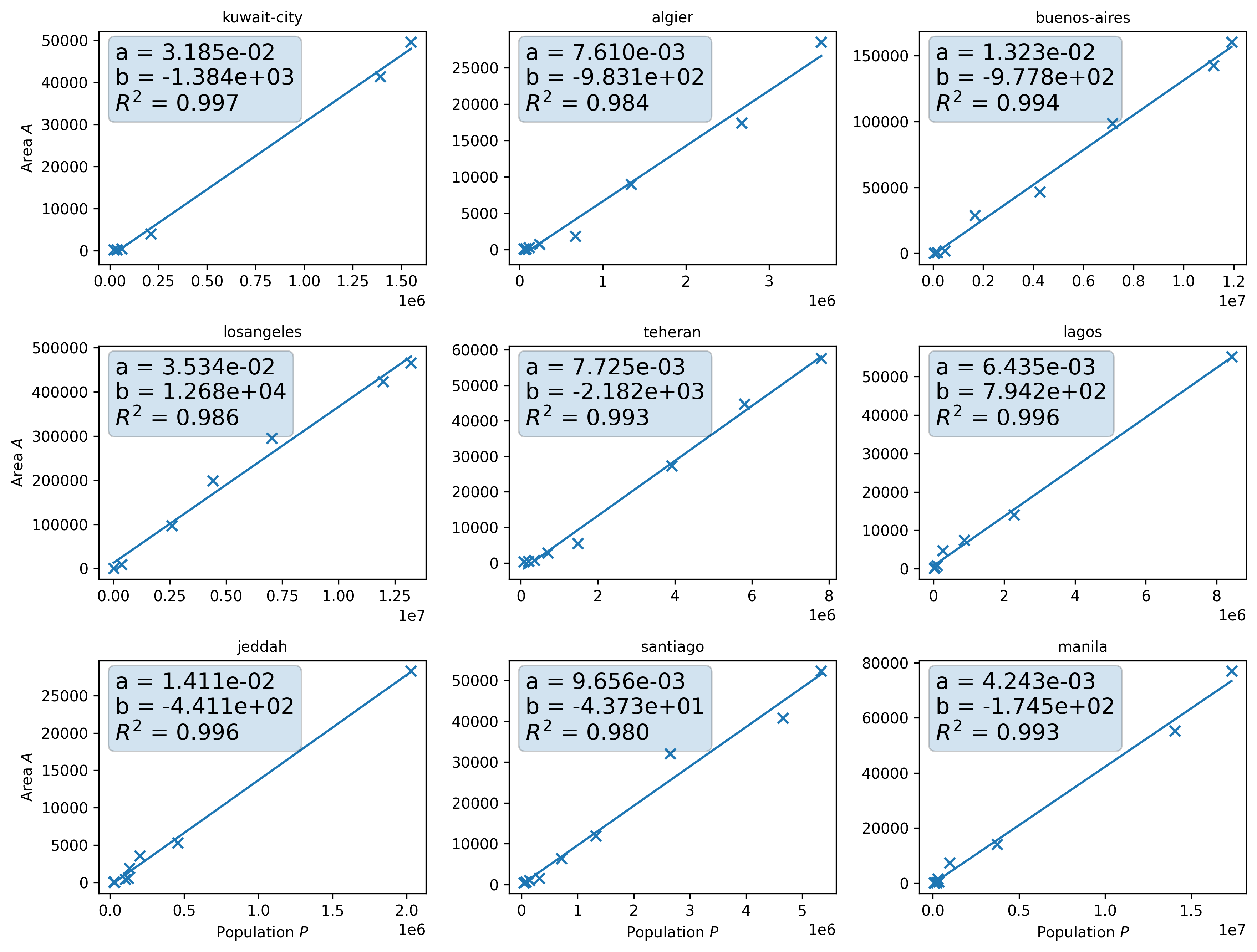}
    \caption{{\bf Examples of longitudinal linear dependency between area ($A$) and population ($P$).} a) For four cities of the WSFEvo dataset~\cite{wsfevo} studied in \cite{marquis2025universal} and b) for nine cities in the~\cite{angel2012atlas} dataset. On each panel, the points represent the data (annual for the WSFEvo dataset, sporadic for the~\cite{angel2012atlas} dataset). For each city, the inverse density ($a$), the intercept ($b$) and the goodness-of-fit ($R^2$), estimated by ordinary least-square regression, are reported for each city. The line on each panel represent the best linear fit~$A = a P + b$.}
    \label{fig:linear}
\end{figure}

\begin{figure}
    \centering
    \includegraphics[width=0.9\linewidth]{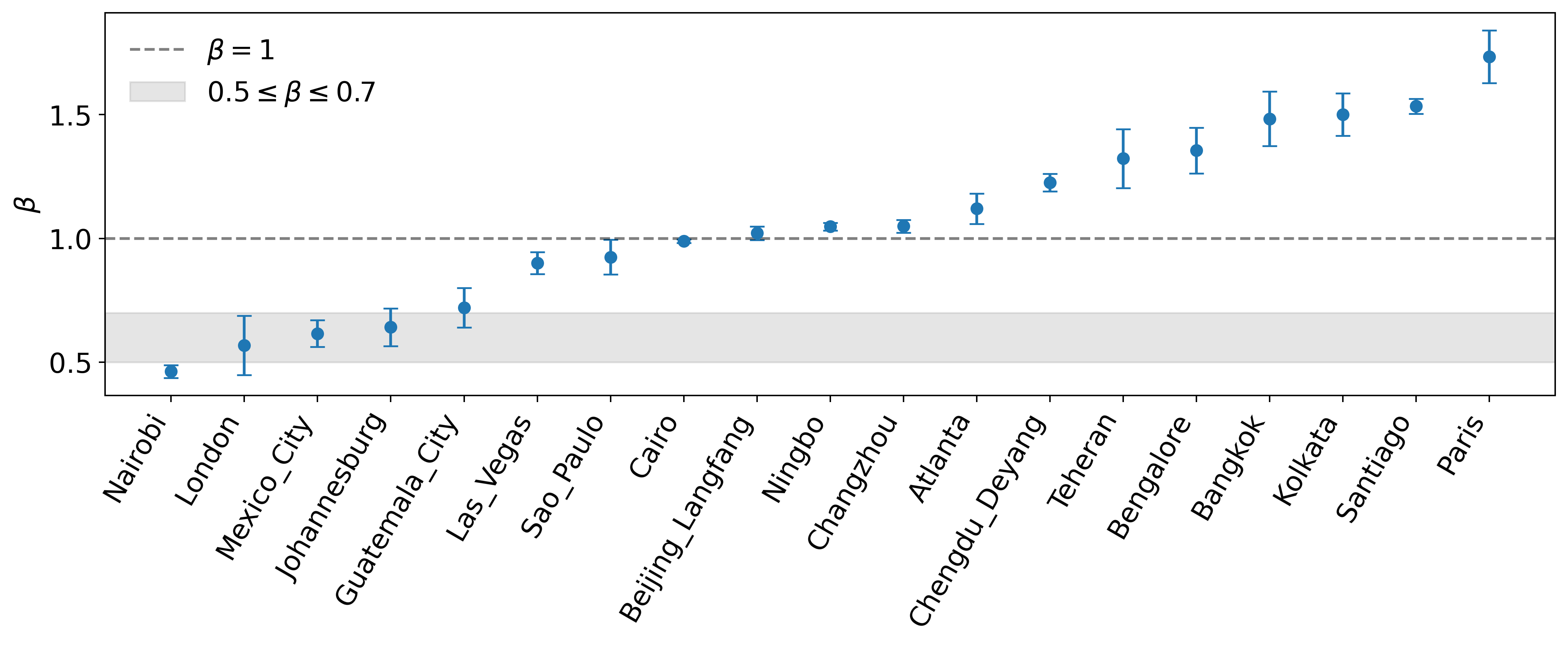}
    \caption{Measured power-law exponents for the cities in the dataset of~\cite{marquis2025universal}. The exponents and error bars were computed using an ordinary least-square regression in log-log coordinates. A dashed line represents the linear case and the greyed are the measured values for the transversal exponent.}
    \label{fig:longitudinal_powerlaw}
\end{figure}

In Fig.~\ref{fig:linear} are shown examples of linear fits for the area-population curves in the case the recent data of~\cite{wsfevo} (left panel) and historical data (1800-2000, right panel). The coefficients for the linear fits and the goodness-of-fit are shown in the legend. In the case of the data~\cite{wsfevo}, we also did longitudinal power law fits and the results are displayed in Fig.~\ref{fig:longitudinal_powerlaw}. The results range between~$0.5$ and~$1.8$, and therefore in general, do not agree with the transversal scaling~$\beta_T \approx 0.5-0.7$.

\subsection{Wages: longitudinal exponent}

\begin{figure}
    \centering
    \includegraphics[width=0.6\linewidth]{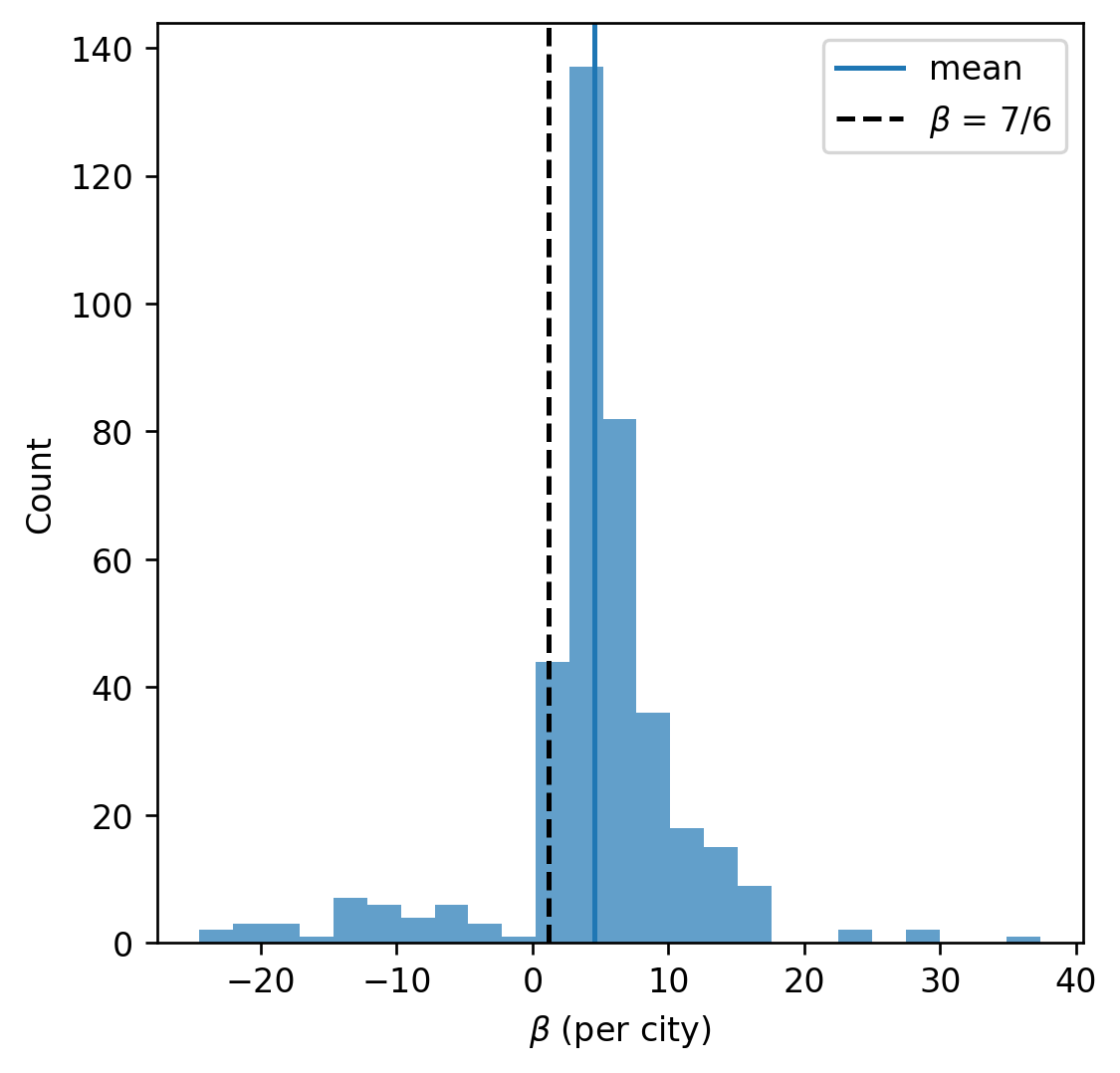}
    \caption{Distribution of individual longitudinal exponents $\beta_i$ obtained from power-law fits for each city. The vertical line indicates the transversal exponent $\beta_T$ estimated from cross-sectional data. The broad distribution highlights the strong heterogeneity of individual city dynamics.}
    \label{fig:betai_dist}
\end{figure}

We show in Fig.~\ref{fig:betai_dist} the distribution of the individual longitudinal exponents $\beta_i$, obtained by fitting a power-law relation for each city. The distribution exhibits a broad dispersion around its mean value, revealing a significant heterogeneity in the longitudinal dynamics across cities. 

For comparison, we also indicate on the same figure the mean and the theoretical value of the transversal exponent $\beta_T$ (vertical line). Strikingly, this theoretical value does not correspond to any typical individual behavior, but rather lies within a wide distribution of exponents, further emphasizing that the transversal exponent does not reflect the dynamics of a representative city.

\subsection{Wages: decomposition of the transversal exponent}

\begin{figure}
    \centering
    \includegraphics[width=1\linewidth]{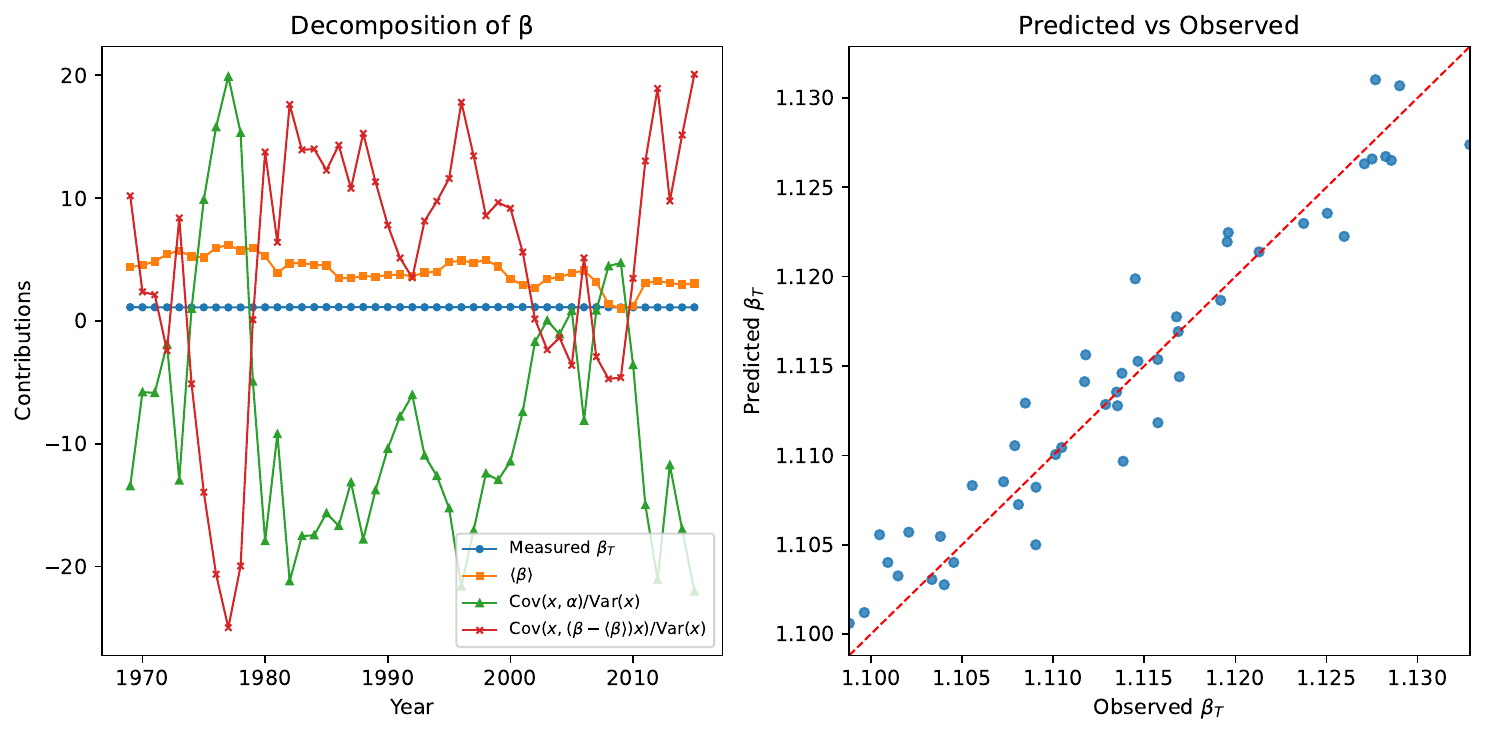}
    \caption{{\bf Wages: scaling decomposition.}(Left) Decomposition of the contributions to the transversal exponent, discussed in Sec.~\ref{subsec:general}. It is important to note the large deviation between the average longitudinal exponent and the transversal exponent. (Right) Comparison between predicted and measured transversal exponents. Red line: identity.}
    \label{fig:wages_decomposition}
\end{figure}

We display in Fig.~\ref{fig:wages_decomposition} each term present in the equation (Eq.~7 in the main text)
\begin{align} \label{eq:decomp}
\beta_T &=\langle\beta\rangle
+ \frac{\mathrm{Cov}(x,\alpha)}{\mathrm{Var}(x)} + \frac{\mathrm{Cov}(x, (\beta-\langle\beta\rangle) x)}{\mathrm{Var}(x)} \,.
\end{align}
for the wages data discussed in the main text. While the transversal exponent (blue points) is stable, it is tendentially much smaller than the estimated average longitudinal exponent~$\langle \beta \rangle$ (orange points) which is typically in the range~$4-5$ but varies noticeably. Remarkably, the only period in which these exponents coincide very well is just afterwards the 2008 crisis. The other correlation terms (green and red points) have approximately opposite trends and their variations largely cancel out. The agreement between predicted and measured transversal exponent (right panel) is robust, even though some fluctuations are clearly noticeable. A reasonable explanation for this phenomena is the large variation of single longitudinal exponent~$\beta_i$ around~$\langle \beta \rangle$.

\subsection{Congestion delay}

We consider here the case of congestion-induced delays, as analyzed in \cite{depersin2018global}, using a dataset of 101 US cities over the period 1982--2014 (for details on the dataset, see \cite{tti2015mobility}). This dataset is particularly relevant, as it is known to exhibit strong heterogeneity in the temporal evolution of individual cities, together with a robust superlinear scaling at the cross-sectional level.

\begin{figure}[htbp!]
  \centering
  \includegraphics[width=0.5\textwidth]{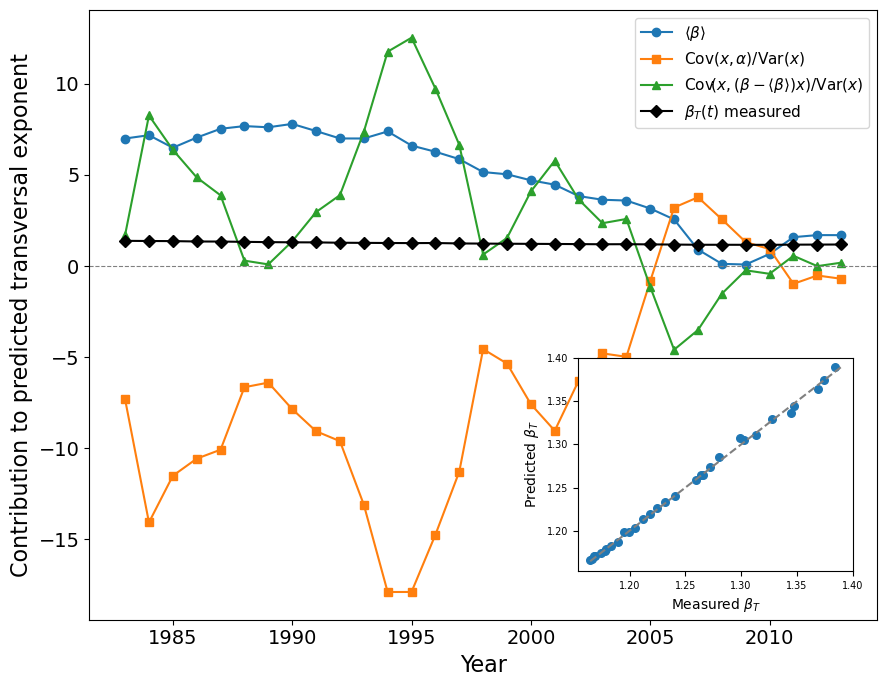}
  \caption{Decomposition of the transversal exponent into its different contributions, as described by Eq.~\eqref{eq:decomp}. Inset: Comparison between the transversal exponent measured directly from the data and the value estimated using Eq.~\eqref{eq:decomp}. Data from \cite{depersin2018global}.}
  \label{fig:congestion}
\end{figure}

In Fig.~\ref{fig:congestion}(inset), we compare the transversal exponent $\beta_T$ directly measured from the data with the value predicted by 
\begin{align} \label{eq:decomp}
\beta_T &=\langle\beta\rangle
+ \frac{\mathrm{Cov}(x,\alpha)}{\mathrm{Var}(x)} + \frac{\mathrm{Cov}(x, (\beta-\langle\beta\rangle) x)}{\mathrm{Var}(x)} \,.
\end{align}
As for wages, we observe an excellent agreement between the two, confirming that the covariance-based decomposition provides an accurate description of the origin of the transversal scaling.

In Fig.~\ref{fig:congestion}, we display the contribution of the different terms entering Eq.~\eqref{eq:decomp}. We observe a behavior similar to that found for the area--population relation. In particular, the average longitudinal exponent $\langle \beta \rangle$ differs significantly from the measured transversal exponent $\beta_T$, which varies only weakly in the range $[1.31, 1.39]$.

This discrepancy is explained by the covariance terms, whose contributions are large but partially compensate each other, resulting in a relatively small net correction. In this case, the combined effect of these terms is negative, leading to a transversal exponent that is smaller than the average longitudinal exponent and remains remarkably stable over time.

At the level of individual cities, the dynamics of congestion delays are highly heterogeneous, as already emphasized in \cite{depersin2018global}, with large temporal fluctuations and, in some cases, very large effective exponents when fitted independently. In contrast, the cross-sectional analysis yields a smooth and robust superlinear scaling, reflecting the overall increase of congestion delays with population.

However, this apparent regularity is misleading: the value of the transversal exponent $\beta_T$ does not directly reflect the intrinsic dynamics of individual cities. Instead, it emerges from the interplay between heterogeneous longitudinal behaviors and statistical correlations across cities. As a consequence, the superlinear scaling observed at the transversal level should not be interpreted as evidence of a universal underlying mechanism governing congestion growth, but rather as a statistical aggregate of diverse and city-specific dynamics.

\bibliography{final/ref_scaling}
\bibliographystyle{IEEEtran}